\documentclass[conference]{IEEEtran}
\IEEEoverridecommandlockouts

\usepackage{tcolorbox}
\tcbuselibrary{breakable}
\usepackage{tabularx}
\usepackage{booktabs}
\usepackage{caption}
\usepackage{xcolor}
\usepackage{graphicx}
\usepackage{url}
\usepackage{array}
\usepackage{paralist}
\usepackage[hidelinks]{hyperref}
\usepackage{orcidlink}
\usepackage{tikz}
\usetikzlibrary{positioning,arrows.meta,fit,backgrounds,calc}

\definecolor{genfill}{HTML}{D6E6F5}
\definecolor{genline}{HTML}{2C6CA8}
\definecolor{extline}{HTML}{6B7785}
\definecolor{opalfill}{HTML}{FBE3C4}
\definecolor{opalline}{HTML}{B8690F}
\definecolor{zappl}{HTML}{F2F3F5}
\definecolor{zgen}{HTML}{EDF3FA}
\definecolor{zfro}{HTML}{F2F3F5}
\definecolor{arrowc}{HTML}{39424B}
\setdefaultleftmargin{1.2em}{1.0em}{}{}{}{} 
\makeatletter
\newcommand{\linebreakand}{%
  \end{@IEEEauthorhalign}
  \hfill\mbox{}\par
  \mbox{}\hfill\begin{@IEEEauthorhalign}
}
\makeatother

\newcommand{\rih}[1]{\vspace{0.6ex}\noindent\textbf{#1}}

\newcommand{\dialogbox}[2]{%
  \begin{center}\footnotesize
  \setlength{\fboxsep}{4pt}%
  \fcolorbox{black!45}{black!5}{\begin{minipage}{0.93\columnwidth}
  \textbf{#1}\par\vspace{0.5ex}\raggedright #2
  \end{minipage}}\end{center}}

\begin{document}

% \title{Genesis Platform Capabilities for Cross-Facility\\ Autonomous Experiments:
% The OPAL Plant Phenotyping Use Case}

\title{Exploring the Genesis Platform Capabilities to Accelerate Scientific Discovery in OPAL}

\author{
    \IEEEauthorblockN{Daniel Rosendo\,\orcidlink{0000-0003-1175-8426}}
    \and
    \IEEEauthorblockN{Renan Souza\,\orcidlink{0000-0002-1794-808X}}
    \and
    \IEEEauthorblockN{Kelsey Carter\,\orcidlink{0000-0001-8327-6413}}
    \and
    \IEEEauthorblockN{John Lagergren\,\orcidlink{0000-0002-8092-7433}}
    \linebreakand
    \IEEEauthorblockN{Frédéric Suter\,\orcidlink{0000-0003-1902-1955}}
    \and
    \IEEEauthorblockN{Shelaine L. Curd\,\orcidlink{0009-0004-6419-7479}}
    \and
    % \IEEEauthorblockN{Gerald A. Tuskan\,\orcidlink{0000-0003-0106-1289}}
    % \and
    \IEEEauthorblockN{David Weston\,\orcidlink{0000-0002-4794-9913}}
    \and
    \IEEEauthorblockN{Rafael Ferreira da Silva\,\orcidlink{0000-0002-1720-0928}}
    \linebreakand
    \IEEEauthorblockA{
       Oak Ridge National Laboratory, Oak Ridge, TN, USA \\
       \{rosendod, souzar, carterkr, lagergrenjh, suterf, curdsl, westondj, silvarf\}@ornl.gov
    }
\thanks{Notice: This manuscript has been authored in part by UT-Battelle, LLC under Contract
No. DE-AC05-00OR22725 with the U.S. Department of Energy. The United States Government
retains and the publisher, by accepting the article for publication, acknowledges that the
United States Government retains a non-exclusive, paid-up, irrevocable, world-wide license
to publish or reproduce the published form of this manuscript, or allow others to do so, for
United States Government purposes. The Department of Energy will provide public access to
these results of federally sponsored research in accordance with the DOE Public Access Plan
(http://energy.gov/downloads/doe-public-access-plan).}}

\maketitle

\begin{abstract}
Autonomous, cross-facility science requires capabilities that no individual project should have to build for itself: managed execution for long-lived services, versioned distribution of models to remote compute systems, governed access to large language models, a shared substrate for experimental data, and end-to-end provenance. The U.S. Department of Energy Genesis Mission platform, delivered through the American Science Cloud, provides these as reusable services. This paper reports how the Genesis platform enables cross-facility experiments and accelerates scientific discovery. We explore the plant phenotyping workflow of the Orchestrated
Platform for Autonomous Laboratories as the exemplar: it couples Oak Ridge National Laboratory's Advanced Plant Phenotyping Laboratory with the Frontier supercomputer. In a 40-day nickel-treatment campaign, the resulting workflow replaced roughly twelve hours of manual analysis with interactive queries returning in seconds to minutes.
\end{abstract}


\IEEEpeerreviewmaketitle

\section{Introduction}

Scientific workflows increasingly span facilities. An experiment begins at an instrument, its data are analyzed on a leadership-class supercomputer, and the results are interpreted through services running in the cloud~\cite{grassroots,antypas,superfacility}. Building such a workflow has historically meant assembling its supporting infrastructure from scratch: somewhere to run long-lived services, a way to get a trained model onto the compute system, credentials and quota for language model access, a place to publish derived data, and some means of recording what happened. Each project solves these problems separately, and each solution is difficult to reuse.

The U.S. Department of Energy (DOE) Genesis Mission addresses this by treating these concerns as \emph{platform capabilities}. Delivered through the American Science Cloud (AmSC), they are operated as shared services with stable interfaces, so that a project integrates against a capability rather than reimplementing it. This paper presents what that shift makes possible in practice. We report on five capabilities, namely a managed cloud Kubernetes execution environment, an MLflow model catalog, the Model Access Gateway (MAG) LLM inference service, a data lakehouse, and the Flowcept provenance service, and show that together they supply most of the non-domain-specific machinery a cross-facility agentic experiment requires.

Our exemplar is the plant phenotyping workflow of the Orchestrated Platform for Autonomous Laboratories (OPAL). It is a demanding use case because it exercises every capability at once and spans three environments: Oak Ridge National Laboratory (ORNL)'s Advanced Plant Phenotyping Laboratory (APPL) acquires imagery continuously, the Frontier supercomputer performs Vision Transformer (ViT) inference and trait extraction, and AI agents hosted on the platform hold a natural language conversation with biologists. We stress at the outset that OPAL is a use case, not the contribution. The capabilities described here are domain-agnostic, and Section~\ref{sec:general} separates what OPAL inherited from the Genesis platform from what it had to build itself.

% Equally important is what the platform does \emph{not} provide yet. Bulk data movement, secure cross-facility messaging, agent hosting on HPC systems, and scheduler-level parallelism are supplied by external services, namely Globus, S3M, Academy, and Parsl. This division is deliberate, and getting it right is part of the design: a platform that absorbs these would duplicate mature software and would have to follow each facility's operational policy. Specifically, this paper makes the following contributions:

Equally important is what the platform does \emph{not} provide yet. Bulk data movement, secure cross-facility messaging, and agent hosting on HPC systems. This division is deliberate, and getting it right is part of the design: a platform that absorbs these would duplicate mature software and would have to follow each facility's operational policy. Specifically, this paper makes the following contributions:

\begin{compactenum}
\item We characterize the requirements that cross-facility autonomous experiments impose on
a science platform, and distinguish platform concerns from facility and domain concerns
(Section~\ref{sec:background}).
\item We describe five Genesis Platform capabilities and the interface each exposes to a
use case (Section~\ref{sec:capabilities}). Additionally, we show how they compose with external services the platform does not subsume, and give the resulting division of responsibility.
\item We demonstrate the composition end to end through the OPAL workflows and a production
deployment over a 40-day campaign, reporting latency and provenance overhead
(Sections~\ref{sec:workflows} and~\ref{sec:eval}).
\item We generalize from the OPAL integration, stating what new use cases inherit and
what they must supply (Section~\ref{sec:general}).
\end{compactenum}

\section{Background and Requirements}
\label{sec:background}

\rih{The plant phenotyping use case.} APPL is a DOE facility at ORNL that exploits natural genetic diversity to improve plant resilience. Its conveyor moves up to 520 trays along a 700-foot track through five aboveground imaging stations, operating 24 hours a day and imaging as many as 10,400 plants over a multi-week campaign across eight modality groups: RGB, hyperspectral, thermal, multispectral, chlorophyll fluorescence, 3D laser, and below-ground RGB and near-infrared. The facility generates on the order of hundreds of gigabytes per day. While acquisition is fully automated, converting imagery into biologically meaningful traits was not. In a representative 40-day campaign that step consumed roughly two hours of a scientist's time per day across six days of analysis, and results arrived only after the campaign had ended. This is the phenotyping bottleneck~\cite{furbank} in modern form: the constraint has migrated from the instrument to the analyst.

\rih{Why this is a platform problem.} Closing that gap requires more than a better script. The trained segmentation model lives where it was trained but must execute on Frontier. Imagery must move from the greenhouse to an analysis server and then to the supercomputer's file system. Derived traits must be published somewhere the interpretation layer can reach. A language model must be reachable under institutional policy. And because non-deterministic model participate in the analysis, the system must record what was asked, what was decided, what code ran, on which hardware, and over which inputs, in a form that can be queried after the fact.

None of these are plant phenotyping problems. They recur in any workflow that couples an
instrument, an HPC system, and an interpretation layer, which is precisely the argument for solving them once at the platform level.

\rih{Requirements.} From this we expose five requirements on a science platform. First, \emph{managed execution}: use cases need somewhere to run long-lived, networked services without operating infrastructure themselves. Second, \emph{model portability}: a model trained or registered in one environment must be retrievable, by version, from a compute system elsewhere. Third, \emph{governed model access}: language model inference must be available through a single controlled endpoint rather than through per-project credentials. Fourth, \emph{a shared data substrate}: experimental metadata and derived products must be publishable and queryable across facility boundaries. Fifth, \emph{provenance}: agent decisions, model exchanges, and computational steps must be captured in a queryable form. Two further requirements, namely bulk data movement and secure cross-facility messaging, are real but are met by external services rather than by the platform itself.%, for reasons we return to in Section~\ref{sec:external}.

\begin{figure*}[!t]
\centering
\begin{tikzpicture}[font=\scriptsize, >=Latex]
\tikzset{
  gen/.style={draw=genline, rounded corners=1.5pt, align=center, inner sep=2.5pt,
              minimum height=6mm, text width=21mm, fill=genfill},
  ext/.style={draw=extline, dashed, rounded corners=1.5pt, align=center, inner sep=2.5pt,
              minimum height=6mm, text width=21mm, fill=white},
  opal/.style={draw=opalline, very thick, rounded corners=1.5pt, align=center,
               inner sep=2.5pt, minimum height=6mm, text width=21mm, fill=opalfill},
  lbl/.style={font=\scriptsize\bfseries, black!80},
  fl/.style={->, thick, arrowc},
  fl2/.style={<->, thick, arrowc},
}
% zone backgrounds
\fill[zappl, rounded corners=3pt] (0.00,0.60) rectangle (2.95,4.35);
\fill[zgen, rounded corners=3pt] (3.30,0.60) rectangle (10.75,4.35);
\fill[zfro, rounded corners=3pt] (12.10,0.60) rectangle (17.60,4.35);
\node[lbl] at (1.47,4.10) {ORNL's APPL};
\node[lbl] at (7.30,4.10) {Genesis Platform};
\node[lbl] at (14.85,4.10) {OLCF Frontier};

% biologist: the human user, deliberately unboxed so that it is not read as one of
% the three component categories. The chat UI is the entry point.
\begin{scope}[shift={(4.25,4.95)}, draw=arrowc, line width=0.6pt]
  \draw (0,0.20) circle (0.14);
  \draw (0,0.06) -- (0,-0.22);
  \draw (-0.20,-0.04) -- (0.20,-0.04);
  \draw (0,-0.22) -- (-0.16,-0.48);
  \draw (0,-0.22) -- (0.16,-0.48);
\end{scope}
\node[font=\scriptsize, black!85, anchor=west] at (4.55,4.95) {\textbf{Plant biologist}};

% APPL facility
\node[opal, text width=20mm] (green) at (1.47,3.35) {Greenhouse\\(imaging)};
\node[opal, text width=20mm] (aserv) at (1.47,2.05) {Analysis Server};
\draw[fl] (green) -- node[right, font=\tiny, xshift=1pt] {Globus} (aserv);

% EKS cluster with the OPAL agent stack
\draw[draw=genline, fill=genfill!70, rounded corners=3pt] (3.45,2.45) rectangle (7.68,3.95);
% label moved to the top-right corner so it does not collide with the biologist arrow
\node[font=\tiny\bfseries, genline, anchor=east] at (7.61,3.78) {EKS cluster};
\node[opal, text width=10mm] (ui) at (4.25,3.10) {Chat UI};
\node[opal, text width=19mm] (cop) at (6.55,3.10) {Co-Scientist Agent (Academy)};

% platform services
\node[gen, text width=18mm] (lake) at (4.35,1.35) {Data lakehouse};
\node[gen, text width=22mm] (mag) at (6.75,1.35) {MAG LLM inference};
\node[gen, text width=22mm] (mlf) at (9.35,1.35) {Model catalog};

% exchange (external)
\node[ext, rotate=90, text width=15mm, minimum height=5mm] (s3m) at (11.42,3.10)
  {S3M Redis};

% Frontier
\node[opal, text width=24mm] (cag) at (13.85,3.35) {Compute Agent (Academy)};
\node[ext, text width=13mm] (parsl) at (13.35,2.15) {Parsl};
\node[opal, text width=20mm] (vit) at (16.25,2.15) {ViT inference (GPU nodes)};
\node[ext, text width=26mm] (orion) at (14.90,1.15) {Orion file system};

% provenance bar (platform capability)
\node[draw=genline, rounded corners=1.5pt, fill=genfill, minimum height=5mm,
      text width=170mm, align=center] at (8.80,-0.30)
  {\textbf{Flowcept provenance service} --- agent decisions, dialog interactions, LLM
   exchanges, data products, and HPC job metadata};

% flows
\draw[fl2] (4.25,4.42) -- (ui.north);
\draw[fl2] (ui.east) -- (cop.west);
\draw[fl2] (5.75,2.82) -- (lake.north);
\draw[fl] (6.95,2.82) -- (mag.north);
\draw[fl2] (cop.east) -- (11.00,3.10);
\draw[fl2] (11.84,3.10) -- (cag.west);
\draw[fl] (cag.south) -- (parsl.north);
\draw[fl] (parsl.east) -- (vit.west);
\draw[fl] (vit.south) -- (16.25,1.48);
\draw[fl] (mlf.east) -- node[above, font=\tiny] {ViT checkpoint} (13.30,1.35);
% Globus long-haul transfer
\draw[fl] (aserv.south) -- (1.47,0.22) -- node[above, font=\tiny, pos=0.62]
  {Globus transfer: AI-ready arrays} (14.90,0.22) -- (orion.south);
\end{tikzpicture}
\caption{The OPAL workflow composed from Genesis platform capabilities and external
services. The biologist interacts with OPAL through the chat UI, which serves as the human-facing entry point to the Co-Scientist Agent. Blue boxes are Genesis platform capabilities; white boxes with dashed borders are external services the platform does not subsume; amber boxes with thick borders are OPAL use case components. Only the amber boxes are specific to plant phenotyping.}
\label{fig:arch}
\end{figure*}
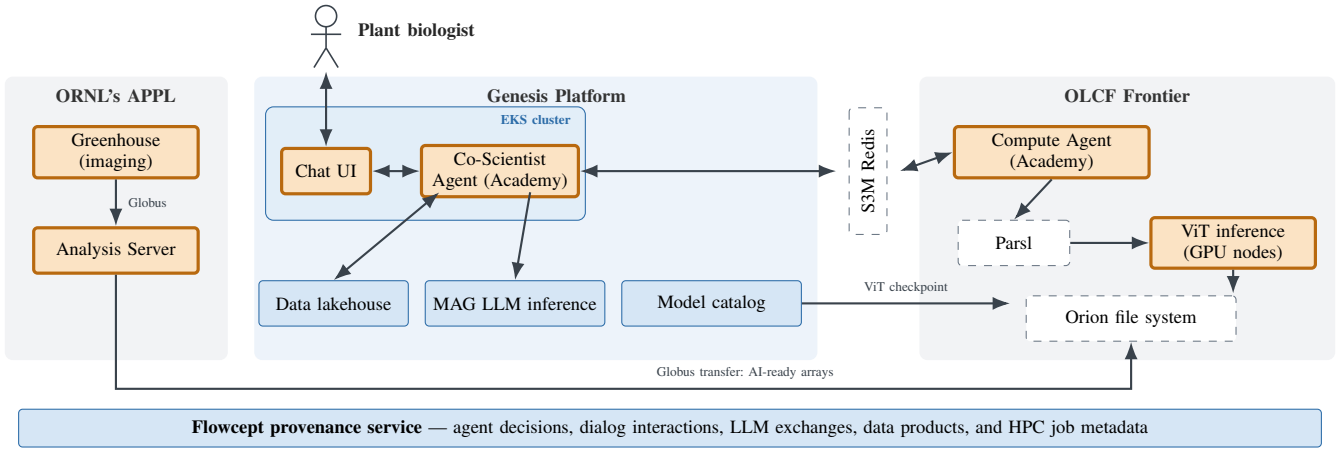

\section{Genesis Platform \& Facility-operated Capabilities Explored in OPAL}
\label{sec:capabilities}

Table~\ref{tab:stack} shows the division of service responsibility across a  typical OPAL workflow  while Fig.~\ref{fig:arch} shows the composition of these services.
The shape of the workflow derives from the capabilities. Globus moves imagery data from the greenhouse to the APPL analysis server, where it is preprocessed into AI-ready arrays, and then to Frontier's Orion file system. On Frontier, a compute agent hosted by Academy dispatches vision transformer inference through Parsl, retrieving the model from the platform's MLflow catalog when no local checkpoint is present. Derived products are published to the data lakehouse. On the platform side, a conversational agent runs in the cloud Kubernetes cluster, queries the lakehouse, and reaches language models through MAG. The two agents exchange control messages over the S3M Redis service. Throughout, Flowcept records what happened.

\begin{table}[!h]
\renewcommand{\arraystretch}{1.15}
\caption{Division of responsibility across the workflow}
\label{tab:stack}
\centering
\footnotesize
\begin{tabular}{|p{0.245\columnwidth}|p{0.6\columnwidth}|}
\hline
\multicolumn{2}{|l|}{\textbf{Genesis platform capabilities}} \\
\hline
Cloud Kubernetes cluster & Hosts the OPAL chat UI and the Co-Scientist agent \\
\hline
Model catalog & MLflow; stores versioned ViT weights, pulled by the compute agent on
Frontier \\
\hline
MAG LiteLLM & Model Access Gateway; serves all LLM inference for the agent \\
\hline
Data lakehouse & Stores experiment metadata and AI-ready data: thumbnails, masks, feature
files \\
\hline
Flowcept & Captures and serves provenance across both facilities \\
\hline
\multicolumn{2}{|l|}{\textbf{External services and frameworks}} \\
\hline
Globus & Moves imagery\hfill\hfill\linebreak greenhouse\,$\rightarrow$\,analysis server\,$\rightarrow$\,Frontier \\
\hline
S3M Redis & Secure cross-facility agent-to-agent messaging \\
\hline
Academy & Deploys agents across federated resources\hfill\hfill\linebreak (e.g., Kubernetes cluster and Frontier) \\
\hline
Parsl & Scales ViT inference across Frontier GPUs \\
\hline
\multicolumn{2}{|l|}{\textbf{Supplied by the use case}} \\
\hline
OPAL components & APPL analysis server, Co-Scientist agent, compute agent logic, ViT model, trait extraction \\
\hline
\end{tabular}
\end{table}

\subsection{Genesis Platform Capabilities}

\rih{Managed execution environment.} The platform operates a cloud-hosted Kubernetes cluster on which use cases deploy their services. OPAL runs its conversational agent there. The Genesis platform added value is not that Kubernetes is novel but that the cluster is operated, authenticated, and network-connected by the platform: a use case obtains a place for long-lived, externally reachable services without negotiating hosting, and services deployed there sit inside the same trust boundary as the other capabilities, so reaching the model catalog, gateway, or data lakehouse does not require separate credentials for each.

\rih{Model catalog.} The catalog is an MLflow deployment that holds versioned model artifacts. Its role is to decouple where a model is produced from where it executes. OPAL uploads its fine-tuned vision transformer weights to the catalog; when the compute agent runs on Frontier it first looks for a local checkpoint and, finding none, downloads the model from the catalog. This cache-with-fallback pattern is worth noting because it is the common case for HPC inference: the catalog is consulted rarely, on a cold node or after a model update, while steady-state execution reads from the file system. It also makes the model version an explicit, recorded property of a run rather than an implicit property of whatever file happened to be on disk, which is what allows provenance to answer which weights produced a given result.

\rih{Model Access Gateway.} MAG~\cite{mag} is a LiteLLM-based inference service that exposes language models through a single endpoint. The OPAL Co-Scientist agent performs all of its inference through it, whether classifying intent, generating analysis code, or writing a report. The gateway centralizes credentials and quota, lets the underlying model be changed without touching agent code, and provides one place where prompts and responses can be observed for provenance. That last property matters more than it first appears: because every model invocation crosses a single boundary, provenance coverage of the reasoning layer is a property of the architecture rather than of programmer discipline.

\rih{Data lakehouse.} The lakehouse is the shared substrate between the facility side and the interpretation side of the workflow. OPAL uploads experiment metadata together with AI-ready data, including plant thumbnails, segmentation masks, and feature files, and the Co-Scientist agent queries it. This is what allows the conversational layer to be hosted on the platform while the data are produced on a supercomputer: publication to the lakehouse is the interface between them, so the agent does not need an account, a file system mount, or a scheduler allocation on Frontier in order to answer a question about the experiment.

\rih{Provenance service.} Flowcept~\cite{flowcept} captures fine-grained provenance across distributed components and curates it into a graph aligned with the W3C PROV model~\cite{provdm} and its agentic extension~\cite{provagent}. It records agent decisions, dialog interactions, LLM exchanges, generated code and its execution outcome, data transformations, and HPC job metadata including node and GPU allocation and queue wait. Because the records are aligned to one model, agentic and conventional steps compose into a single queryable graph spanning both facilities, and because they persist independently of any model's context window, they provide a definitive history rather than a recollection. Users query this history in natural language through a provenance agent that translates questions into queries over the records and summarizes the results, an approach evaluated in prior work~\cite{provllm}. Flowcept is used by OPAL today and is being integrated as the provenance service for the Genesis platform, so that provenance becomes a property use cases inherit rather than one each must deploy for itself.

\subsection{Facility-operated Capabilities}
\label{sec:external}

\rih{S3M Redis for cross-facility messaging.} Agents deployed on the Genesis platform Kubernetes cluster and on Frontier cannot interact  directly. The Secure Scientific Service Mesh (S3M)~\cite{s3m} provides a token-authenticated Redis service spanning ORNL's administrative and network boundaries, which both sides reach outbound, so neither facility exposes a new inbound service. This is necessarily a facility-operated capability. It derives its security posture from the facility whose perimeter it crosses, and a platform-operated substitute would have to re-establish that posture from outside.

\rih{Globus Transfer for data movement.} Globus~\cite{globus} moves imagery data  from the APPL greenhouse to the APPL analysis server and from there to Frontier. Bulk transfer between facility endpoints is a solved problem with mature, widely deployed tooling and facility-specific endpoint configuration; a platform that reimplemented it would duplicate that software and inherit the obligation to track every facility's transfer policy. The platform's concern begins once data are ready to be published, which is where the lakehouse takes over.

\subsection{DOE Community Software}
\label{sec:doecommunity}

\rih{Academy for remote agent deployment.} The compute agent runs on a Frontier login node under Academy~\cite{academy}, a framework for deploying agents across federated research infrastructure. Its actor model, in which agents encapsulate local state and coordinate by message passing, fits a topology in which the APPL Co-Scientist agent runs in the Genesis Platform Kubernetes cluster and the Compute agent on Frontier, and its exchange abstraction let us adopt S3M Redis service as the transport without changing agent code. Frameworks designed for cloud-hosted assistants generally assume a central orchestrator and a cloud-bound message bus, neither of which holds here. No single orchestrator can span both facilities.

\rih{Parsl for scaling on HPC.} Heavy computation is isolated into functions annotated as Parsl apps~\cite{parsl}, which separates interactive agent code from batch execution and delegates scheduler interaction. Two main properties of Parsl fit this setting. First, its task graph is constructed at runtime and returns futures, which suits work whose extent is not known until a plan is confirmed, since the plants, images, and modalities named there determine how many inferences are issued, and which lets the agent remain responsive while they run. And its block-based provisioning packs the many short inference tasks of a targeted job into a single allocation, which matters because cold path latency is dominated by queue wait rather than by inference (Section~\ref{sec:eval}). Second, its executor configuration further isolates what is machine specific, so targeting a system other than Frontier is a change of configuration rather than of agent code.

\dialogbox{}{
The resulting principle is that the \textbf{Genesis platform} supplies what is common across use cases and stable across facilities, while \textbf{facility-\linebreak operated} and \textbf{DOE community software} supplies what is specific to a machine, a perimeter, or a scheduler. Where the two meet, the interface is a published artifact: a model version in the catalog, a dataset in the lakehouse, a message on the exchange, a provenance record.
}

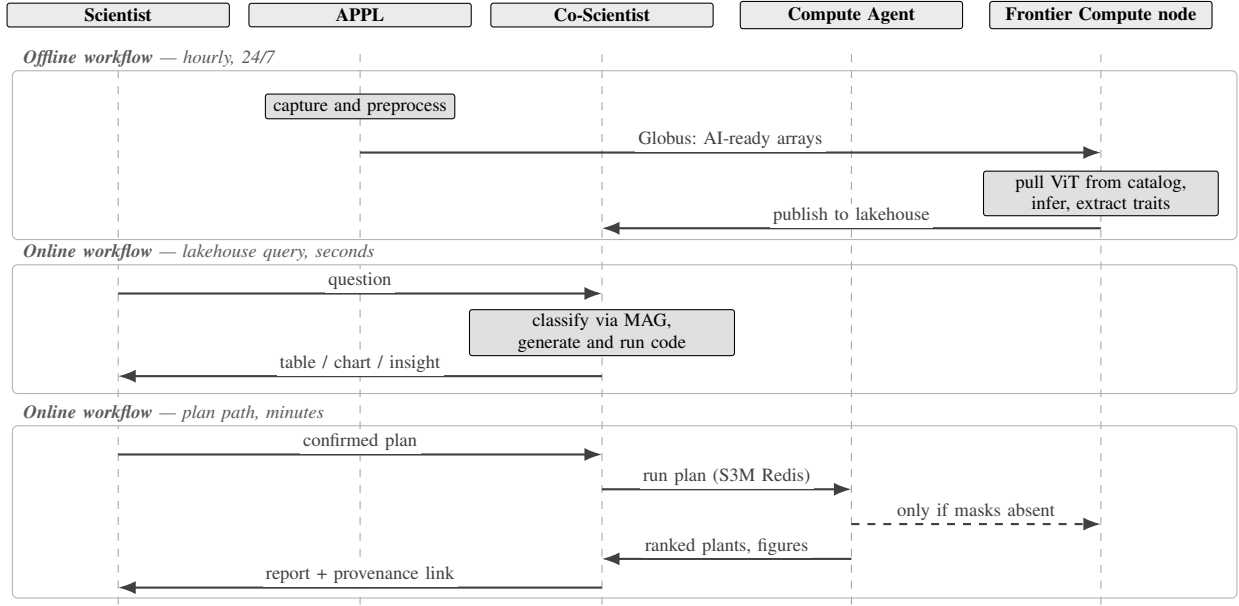
\begin{figure*}[!t]
\centering
\begin{tikzpicture}[font=\small, yscale=1.15, >=Latex]
\tikzset{
  ll/.style={draw, fill=black!8, rounded corners=1pt, align=center, inner sep=2pt,
             text width=28mm, font=\scriptsize\bfseries},
  act/.style={draw, fill=black!12, rounded corners=1pt, align=center, inner sep=1.5pt,
              font=\scriptsize},
  msg/.style={->, thick, black!75},
  ml/.style={above, font=\scriptsize, fill=white, inner sep=1.2pt},
  blk/.style={draw, black!30, rounded corners=2pt},
  ph/.style={font=\scriptsize\itshape, black!65, fill=white, inner sep=1.2pt},
}
% lifelines
\node[ll] (a1) at (1.20,0) {Scientist};
\node[ll] (a2) at (4.40,0) {APPL};
\node[ll] (a3) at (7.60,0) {Co-Scientist};
\node[ll] (a4) at (10.90,0) {Compute Agent};
\node[ll] (a5) at (14.20,0) {Frontier Compute node};
\foreach \x in {1.20,4.40,7.60,10.90,14.20}
  \draw[black!35, dashed] (\x,-0.45) -- (\x,-6.80);

% ---- offline pipeline ----
\node[ph, anchor=west] at (-0.10,-0.50) {\textbf{Offline workflow} --- hourly, 24/7};
\draw[blk] (-0.20,-0.64) rectangle (16.00,-2.58);
\node[act, text width=24mm] at (4.40,-1.05) {capture and preprocess};
\draw[msg] (4.40,-1.58) -- node[ml]{Globus: AI-ready arrays} (14.20,-1.58);
\node[act, text width=30mm] at (14.20,-2.05) {pull ViT from catalog,\\ infer, extract traits};
\draw[msg] (14.20,-2.45) -- node[ml]{publish to lakehouse} (7.60,-2.45);

% ---- online: lakehouse query path ----
\node[ph, anchor=west] at (-0.10,-2.73) {\textbf{Online workflow} --- lakehouse query, seconds};
\draw[blk] (-0.20,-2.87) rectangle (16.00,-4.35);
\draw[msg] (1.20,-3.20) -- node[ml]{question} (7.60,-3.20);
\node[act, text width=34mm] at (7.60,-3.65) {classify via MAG, generate and run code};
\draw[msg] (7.60,-4.15) -- node[ml]{table / chart / insight} (1.20,-4.15);

% ---- online: plan path ----
\node[ph, anchor=west] at (-0.10,-4.58) {\textbf{Online workflow} --- plan path, minutes};
\draw[blk] (-0.20,-4.72) rectangle (16.00,-6.70);
\draw[msg] (1.20,-5.05) -- node[ml]{confirmed plan} (7.60,-5.05);
\draw[msg] (7.60,-5.45) -- node[ml]{run plan (S3M Redis)} (10.90,-5.45);
\draw[msg, dashed] (10.90,-5.85) -- node[ml]{only if masks absent} (14.20,-5.85);
\draw[msg] (10.90,-6.25) -- node[ml]{ranked plants, figures} (7.60,-6.25);
\draw[msg] (7.60,-6.58) -- node[ml]{report + provenance link} (1.20,-6.58);
\end{tikzpicture}
\caption{Simplified sequence diagram of the OPAL workflows. The offline pipeline keeps the lakehouse current; most questions are then answered from it in seconds, and only confirmed plans requiring missing masks reach
the GPU inference stage.}
\label{fig:seq}
\end{figure*}

\section{The OPAL Workflows}
\label{sec:workflows}

OPAL contributes three components of its own: a conversational Co-Scientist agent, the logic of a compute agent, and the domain models and algorithms, namely the fine-tuned vision transformer and the trait extraction routines. These are composed into two main workflows (offline and online), shown in Fig.~\ref{fig:seq}, which meet at the data lakehouse.

\rih{Offline workflow.} Every hour, newly captured images are preprocessed on the APPL analysis server into AI-ready arrays and transferred by Globus to Frontier. A job runs segmentation inference over the new arrays, using a checkpoint pulled from the model catalog if none is cached locally, and trait extraction combines images and masks into per-plant trait profiles. Across a full campaign this amounts to roughly 2,000 inferences per day and hundreds of terabytes of masks, traits, and metadata retained on the parallel file system. Derived products, namely thumbnails, masks, feature files, and harmonized experiment metadata, are published to the lakehouse. The purpose of this pipeline is to make the common case fast: by the time a scientist asks about yesterday's imagery, the traits exist.

\rih{Online workflow.} The Co-Scientist agent classifies each message in three stages.  An explicit prefix allows the agent to select a mode directly. Otherwise, it applies deterministic keyword rules and only if both fail, it consults a model through the Genesis platform's MAG. The chat interface exposes four modes: 1) Greetings return in under a second with zero or one model calls; 2) Knowledge base questions retrieve from a curated index of prior findings and synthesize a cited answer in two calls; 3) Tabular and plot queries translate a question into executable code against lakehouse data, taking two to four calls and three to ten seconds, and 4) Plan mode conducts a dialogue and executes on HPC.

\rih{Answering from published data.} The tabular and plot modes generate code against harmonized feature data. The prompt is enriched with a schema, experiment-specific facts, and trait synonyms, because biologists would reasonably write ``greenness'' rather than ``normalized ratio of color channel medians''. Generated code runs in a constrained environment, and failures enter a repair loop in which the error and offending code are returned to the model for correction. A validation stage then checks that genotype names resolve to real accessions and that rankings are well formed, with failures triggering the same repair path. The executed code is returned alongside the answer, so the analysis can be inspected and reproduced. %This is what makes code generation acceptable in a scientific setting: the system does not ask the biologist to trust generated code, it shows the code and the checks it passed.

\rih{Plan path and availability check.} Questions that cannot be answered from published data enter the plan path. The agent conducts a requirements dialogue until the plan is complete and asks for confirmation, which keeps a human in the loop. On confirmation the plan crosses S3M Redis to the compute agent, which resolves the plants and images involved and checks whether the required masks exist. When they do, which is the common case, analysis proceeds immediately; when they do not, on-demand inference is scheduled for a targeted subset. Results are persisted, so the next request for the same traits finds them published. %The scientist is never asked which case applied.

\rih{Report validation.} A generated report is an interpretation, and the system provides a way to check it against the execution record rather than against the model's recollection. When a scientist asks for a report to be validated, the agent retrieves the provenance card for the workflow that produced it and asks the model to assess whether the conclusions are supported by the recorded inputs, parameters, and outputs. Because the card is assembled from provenance rather than conversation history, the check is grounded in what actually executed. It does not remove the need for domain review, and a model critiquing output from the same model has well-understood blind spots; what it does is convert an unverifiable narrative into one whose claims can be traced to specific execution steps.

\section{Deployment and Evaluation}
\label{sec:eval}

\subsection{Setup}

We report on a production deployment supporting a 40-day APPL campaign studying genotypic variation in response to nickel treatment, a question relevant to DOE interests in critical minerals and in plant resilience on marginal soils. The chat interface and the Co-Scientist agent ran in the Genesis platform's Kubernetes cluster, reaching MAG for inference and the lakehouse for data. The compute agent ran on a Frontier login node under Academy, with Parsl dispatching inference to GPU nodes and model weights pulled from the MLflow catalog on cache miss. Globus moved imagery in two hops from the greenhouse to Frontier, and the S3M Redis service carried agent control messages. Trait extraction covered the RGB, infrared, and fluorescence modalities. Flowcept captured provenance throughout.

\subsection{A Representative Session}
The following exchanges illustrate the interaction styles the system supports, including experimental steering. They are representative of production sessions rather than a verbatim transcript of a single one. Questions requiring computation over imagery enter the plan path, where the agent first establishes what is being asked before consuming HPC resources.

\begin{tcolorbox}[
    colback=green!8,
    colframe=green!35,
    boxrule=0.5pt,
    arc=3pt,
    left=6pt,
    right=6pt,
    top=4pt,
    bottom=4pt
]
\small
\textbf{Biologist:} From April 8, 2026 to May 2, 2026, what are the 5 plants with greatest leaf area?
\end{tcolorbox}

\begin{tcolorbox}[
    colback=gray!10,
    colframe=gray!50,
    boxrule=0.5pt,
    arc=2pt,
    left=5pt,
    right=5pt,
    top=4pt,
    bottom=4pt,
    breakable
]
\small
\textbf{Co-Scientist:} Top-5 plants by leaf area (\texttt{rgb2\_area\_mm2}).

\medskip

\begin{center}
\small
\setlength{\tabcolsep}{3pt}
\begin{tabularx}{\linewidth}{c c X c r}
\toprule
\textbf{Rank} &
\textbf{Plant ID} &
\textbf{Genotype} &
\textbf{Ni treatment} &
\textbf{Leaf area} \\
&
&
&
\textbf{(mg kg$^{-1}$)} &
\textbf{(mm$^2$)} \\
\midrule
1 & 29038 & 9       & 200 & 110.23 \\
2 & 28891 & RIL 407 & 50  & 109.48 \\
3 & 29079 & Ecm29   & 0   & 108.26 \\
4 & 29073 & ISU 210 & 500 & 108.21 \\
5 & 29041 & RIL 407 & 0   & 107.22 \\
\bottomrule
\end{tabularx}
\end{center}

% \begin{center}
%     \includegraphics[width=\linewidth]{leaf-area.png}
% \end{center}

\noindent
\textbf{Biological interpretation}

\begin{itemize}
    \item Leaf area is the primary ``nickel tolerance'' trait defined
    for this experiment (maintaining leaf area under high Ni stress).

    \item The highest-area plant (29038, genotype 9) achieves approximately
    110~mm$^2$ at \textbf{200~mg Ni kg$^{-1}$}, well above the control
    plants, indicating strong tolerance to moderate Ni levels.

    \item \textbf{ISU 210} (plant 29073) retains a large leaf area
    (108~mm$^2$) even at the highest tested dose
    (500~mg Ni kg$^{-1}$), suggesting it may be the most Ni-resilient
    accession among the five.

    \item \textbf{RIL 407} shows the second-largest area but only at
    low-to-moderate Ni (0 and 50~mg kg$^{-1}$); its performance at higher
    Ni is unknown and should be examined.

    \item \textbf{Ecm29} performs best at the control but lacks
    data at higher Ni, so its tolerance cannot be judged from this snapshot.
\end{itemize}

Overall, data point to genotypes 9 and ISU 210 as promising candidates for breeding or further screening for Ni-rich soils, while RIL 407 warrants additional testing at higher concentrations.
\end{tcolorbox}

\begin{tcolorbox}[
    colback=green!8,
    colframe=green!35,
    boxrule=0.5pt,
    arc=3pt,
    left=6pt,
    right=6pt,
    top=4pt,
    bottom=4pt
]
\small
\textbf{Biologist:} Now it's end of business on May 3, 2026. Focus only on today's measurements and on plants with leaf area greater than 40mm2. Then, for each plant, compare each measurement to the previous measurement on the same day. Keep only plants with at least one drop greater than 30\% on the day. Finally, make a line plot of leaf area over time, one line per plant. Exclude QC-failed.
\end{tcolorbox}

\begin{tcolorbox}[
    colback=gray!10,
    colframe=gray!35,
    boxrule=0.5pt,
    arc=6pt,
    left=8pt,
    right=8pt,
    top=4pt,
    bottom=4pt
]
\small
\textbf{Co-Scientist:} 

\begin{center}
    \includegraphics[width=\linewidth]{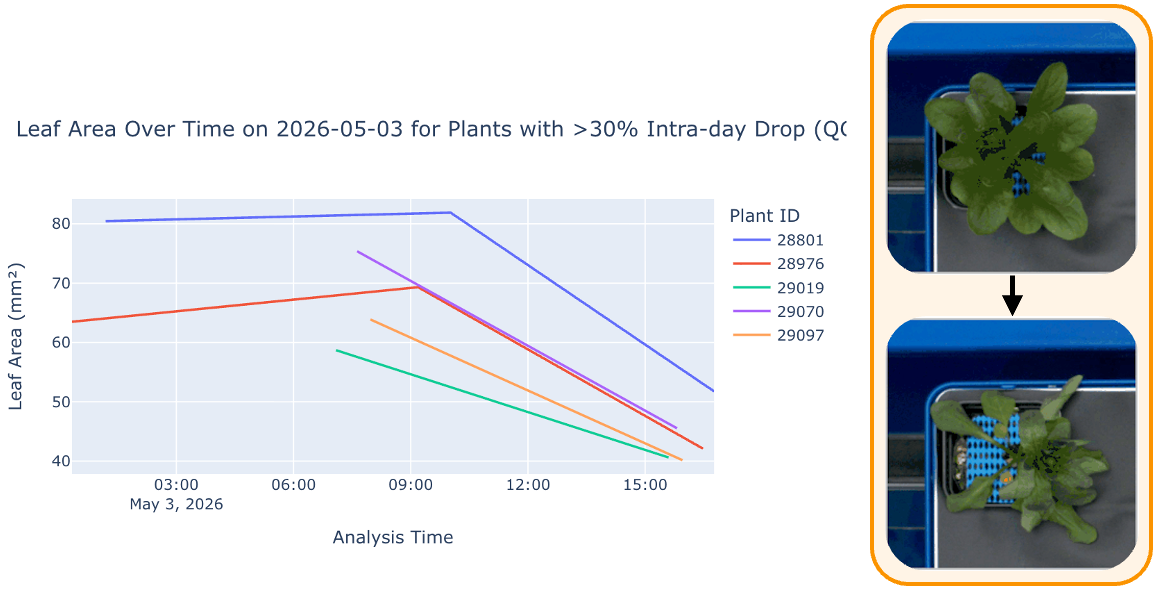}
\end{center}

The \textbf{May 3} plot reveals a systematic mid‑day leaf‑area collapse of greater than 30\% in five plants, most plausibly caused by a temporary watering failure. This procedural stress overrides any Ni‑tolerance signal, so the affected plants should be monitored, the watering system checked, and the experiment repeated under confirmed proper irrigation before using these data for genotype selection.
\end{tcolorbox}

OPAL agents identified a sudden greater than 30\% decrease in plant area and suggested that scientists check water status. After APPL technicians applied the intervention, OPAL agents confirmed plant recovery after rewatering.

\begin{tcolorbox}[
    colback=green!8,
    colframe=green!35,
    boxrule=0.5pt,
    arc=3pt,
    left=6pt,
    right=6pt,
    top=4pt,
    bottom=4pt
]
\small
\textbf{Biologist:} Now it's 11:15am on May 4th, 2026. Yesterday we found that the plants 28801, 28976, 29019, 29070, and 29097 got strong declines in their leaf area caused by bad watering. We watered them as soon as we found this out. Please check their leaf area now. Plot a line plot from yesterday to now with the leaf area of these plants.
\end{tcolorbox}

\begin{tcolorbox}[
    colback=gray!10,
    colframe=gray!35,
    boxrule=0.5pt,
    arc=6pt,
    left=8pt,
    right=8pt,
    top=4pt,
    bottom=4pt
]
\small
\textbf{Co-Scientist:} 

\begin{center}
    \includegraphics[width=\linewidth]{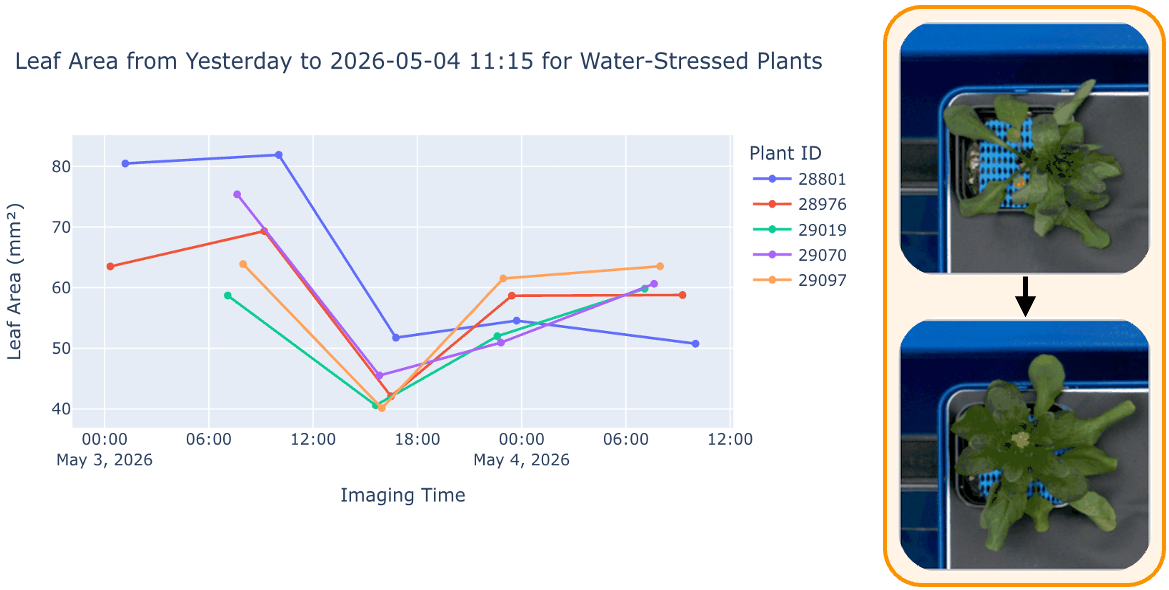}
\end{center}

The plot confirms that the abrupt leaf‑area drops on \textbf{May 3} were caused by the watering failure, and that four of the five plants have already recovered by \textbf{now}, while \textbf{plant 28801} continues to decline and should be watched closely. %Implement the three operational actions above to secure plant health and capture useful recovery metrics.
\end{tcolorbox}

Because every step was recorded, the scientist can interrogate how a result was produced, across both the platform and facility sides of the workflow.

% \begin{tcolorbox}[
%     colback=green!10,
%     colframe=green!10,
%     boxrule=0pt,
%     arc=6pt,
%     left=8pt,
%     right=8pt,
%     top=6pt,
%     bottom=6pt
% ]
% \small
% \textbf{Biologist:} How long did the analysis plan take to run and what was the CPU usage?
% \end{tcolorbox}

% \begin{tcolorbox}[
%     colback=gray!10,
%     colframe=gray!10,
%     boxrule=0pt,
%     arc=6pt,
%     left=8pt,
%     right=8pt,
%     top=6pt,
%     bottom=6pt
% ]
% \small
% \textbf{Co-Scientist:} The analysis plan task took 18.45 seconds to run. The CPU usage during
% execution was as follows: overall CPU utilization changed by 18.45\%,
% user time increased by 0.01 seconds, system time increased by 0.01
% seconds.
% \end{tcolorbox}

\begin{tcolorbox}[
    colback=green!10,
    colframe=green!35,
    boxrule=0.5pt,
    arc=6pt,
    left=8pt,
    right=8pt,
    top=4pt,
    bottom=4pt
]
\small
\textbf{Biologist:} What are the phenotypes of the best plants selected in the last analysis run?
\end{tcolorbox}

\begin{tcolorbox}[
    colback=gray!10,
    colframe=gray!35,
    boxrule=0.5pt,
    arc=6pt,
    left=8pt,
    right=8pt,
    top=4pt,
    bottom=4pt
]
\small
\textbf{Co-Scientist:} The query asks for the phenotypes of the best plants selected in the
last analysis run. The sample result shows \textbf{BESC-856} as the phenotype.
\end{tcolorbox}

\begin{tcolorbox}[
    colback=green!10,
    colframe=green!35,
    boxrule=0.5pt,
    arc=6pt,
    left=8pt,
    right=8pt,
    top=4pt,
    bottom=4pt
]
\small
\textbf{Biologist:} Generate the provenance card.
\end{tcolorbox}

\begin{tcolorbox}[
    colback=gray!8,
    colframe=gray!25,
    boxrule=0.5pt,
    arc=4pt,
    left=7pt,
    right=7pt,
    top=4pt,
    bottom=4pt,
    breakable
]
\small
\textbf{Co-Scientist:}

\smallskip
\textbf{Provenance Card Summary}
\smallskip

\footnotesize

\begin{itemize}
    \item \textbf{Campaign ID:}
    \texttt{9178e5ed-8daf-d4b1487d1c4a}
    \item \textbf{Type:} Pipeline, \textbf{Workflow runs:} 2, \textbf{Total Elapsed (s):} 157.982
\end{itemize}

\smallskip
\small

\textbf{Campaign-level Summary}
\footnotesize

\begin{itemize}
    \item \textbf{Total Workflows:} 2, \textbf{Total Activities:} 4, \textbf{Total Tasks:} 24
    \item \textbf{Status Counts:} \texttt{\{'FINISHED': 24\}}, \textbf{Total Elapsed (s):} 157.982

    \item \textbf{Top Slowest Activities:}
    \scriptsize
    
    \begin{itemize}
        \item \texttt{run\_analysis\_plan}: 7.895s, \texttt{trait\_extraction}: 2.054s
        \item \texttt{dialog\_interaction}: 1.705s, \texttt{run\_segmentation}: 1.071s
    \end{itemize}
\end{itemize}

% ---------------------------------------------------------
\smallskip
\small

\textbf{Workflow Structure}

\begin{center}
\begin{minipage}{\linewidth}
\centering
\ttfamily
\scriptsize

input$\rightarrow$dialog\_interaction$\rightarrow$run\_analysis\_plan\\
$\rightarrow$run\_segmentation$\rightarrow$trait\_extraction$\rightarrow$output
\end{minipage}
\end{center}

% ---------------------------------------------------------
\smallskip
\textbf{Per-workflow Details}

\footnotesize
\medskip
\textbf{Compute Agent Workflow}

\begin{itemize}
    \item \textbf{Workflow ID:}
    \texttt{d0d615f7-b457-8502}, \textbf{Elapsed:} 7.895 s

    \item \textbf{Used:}
    \begin{itemize}
        \scriptsize

        \item \texttt{metadata\_file}: \texttt{/lustre/orion/.../metadata\_file.csv}
        \item \texttt{model}:
    \end{itemize}
\end{itemize}

\begin{tcolorbox}[
    colback=black!88,
    colframe=black!88,
    boxrule=0pt,
    arc=2pt
]
\color{white}
\ttfamily\scriptsize
backbone: vit\_small\_patch8\_224\\
checkpoint\_path: /lustre/orion/\ldots\\
device: cuda\\
id: opal\_vit\_small\_patch8\_224\\
model\_type: vit\\
repository: https://code.ornl.gov/opal/ai\\
sha: 1d1cff06d84973c48e4d859932ac174\\
tile\_size: 448\\
trained\_on: Frontier\\
training\_date: 2025-08-16
\end{tcolorbox}

\begin{itemize}
    \item \textbf{trait extraction package}:
\end{itemize}

\begin{tcolorbox}[
    colback=black!88,
    colframe=black!88,
    boxrule=0pt,
    arc=2pt
]
\color{white}
\ttfamily\scriptsize
modified\_on: 2026-02-03T21:02:03\\
repository: https://code.ornl.gov/opal/ai\\
sha: 1d1cff06d84973c48e4d859932ac174
\end{tcolorbox}

% ---------------------------------------------------------
\smallskip
\small

\textbf{Per-activity Details}

\footnotesize
\smallskip
\textbf{Compute Agent Workflow}

\smallskip
\textbf{\texttt{run\_analysis\_plan}}

\begin{itemize}
    \item \textbf{Used:}
    \begin{itemize}
        \scriptsize
        \item \texttt{experiment\_id}: 41, \texttt{modality}: RGB1
        \item \texttt{performance.formula}: \texttt{1.0*height\_mm}
        \item \texttt{plants.selection}: \texttt{top\_and\_worst}
    \end{itemize}

    \item \textbf{Generated:}
    \begin{itemize}
         \scriptsize
         \item \texttt{best\_plants}: \texttt{[21982]}
        \item \texttt{generated\_masks\_paths}:\\
        \texttt{["41\_\_21982\_\_1966.0\_mok.npy", \ldots]}
        \item \texttt{selected\_plant\_ids}: \texttt{[21982, 21981]}
        \item \texttt{worst\_plants}: \texttt{[21981]}
    \end{itemize}

    \item \textbf{Hosts (tasks per host):}
    \begin{itemize}
         \scriptsize
         \item \texttt{05.frontier.olcf.ornl.gov}: 1 task
    \end{itemize}
\end{itemize}

% ---------------------------------------------------------
\textbf{\texttt{run\_segmentation} (n=15)}

\begin{itemize}
    \item \textbf{Used (aggregated):}
    \begin{itemize}
         \scriptsize
         \item \texttt{image\_path}: 
        type=scalar/categorical; top values=\ldots{} ($\times$15)
    \end{itemize}

    \item \textbf{Generated (aggregated):}
    \begin{itemize}
         \scriptsize
         \item \texttt{mask\_path}: 
        type=scalar/categorical; top values=\ldots{} ($\times$15)
    \end{itemize}

    \item \textbf{Hosts (tasks per host):}
    \begin{itemize}
         \scriptsize
         \item \texttt{05.frontier.olcf.ornl.gov}: 15 tasks
    \end{itemize}
\end{itemize}

\end{tcolorbox}

\subsection{Performance}

Table~\ref{tab:perf} shows the end-to-end latency by interaction class. Two results are worth drawing out. First, the separation between the published and cold paths is large but the published path dominates in practice: because the offline pipeline runs hourly, a scientist asking about recent imagery normally finds the data already in the lakehouse, and the cold path is reached mainly for newly requested modalities or for plants imaged within the last hour. Second, cold path cost is dominated by scheduling rather than by inference. An individual inference takes on the order of ten seconds, and a targeted job of roughly a hundred inferences completes in minutes when GPUs are available; the variance comes from batch queue wait, which in the worst case we observed extends a cold query to tens of minutes or more. This is a property of sharing a leadership-class machine, and it is the main argument for the offline pipeline: the system is fast because it does most of the work before the question is asked.

\begin{table}[!t]
\renewcommand{\arraystretch}{1.15}
\caption{Measured end-to-end latency by interaction class.}
\label{tab:perf}
\centering
\footnotesize
\begin{tabular}{|p{0.60\columnwidth}|p{0.3\columnwidth}|}
\hline
\textbf{Interaction} & \textbf{End-to-end latency} \\
\hline
Greeting / routing only & $<$1\,s \\
\hline
Knowledge base question & 2--5\,s \\
\hline
Query against lakehouse data & 3--10\,s \\
\hline
Plan, masks already available & $<$1\,min \\
\hline
Plan, on-demand inference (8 plants) & $\approx$5\,min \\
\hline
Plan, on-demand inference, contended queue & 30\,min--2\,h \\
\hline
Provenance query & 3--10\,s \\
\hline
\end{tabular}
\end{table}

\rih{Provenance overhead.} Capturing provenance at the granularity described in Section~\ref{sec:capabilities}, covering every agent decision, dialog interaction, model exchange, code execution, and HPC job, added under 1\% to end-to-end runtime. At this granularity and scale, this means that there is no need to trade auditability against performance. Provenance capture can be a default property of the platform rather than a debugging mode enabled after something has gone wrong.

\rih{Comparison with the manual baseline.} The campaign screened 360 pennycress plants for mineral uptake under nickel treatment. The workflow this system replaced consumed roughly two hours of a scientist's time per day across six days of analysis, some twelve hours for the campaign, and it produced a hand-selected subset of traits only after the campaign had concluded. The agentic workflow extracts the full trait panel for every plant at every imaging round and completes within minutes of image acquisition. Measured against that baseline, the system yields on the order of $100\times$ more traits per run and reduces analysis time by roughly $600\times$. The change in kind matters more than either figure. As answers now arrive while the experiment is still running, scientists can adjust the experiment in response to what the most recent data show. Participating biologists project that the agentic system will at least double their productivity on analysis-bound campaigns.

\subsection{Limitations and Failure Modes}

Because the workflow spans a cloud platform, DOE facilities, and external services, its failure modes are distributed, and how it degrades matters as much as how it performs. The behavior is deliberately asymmetric. Failures in the interpretation layer are recoverable and visible. When generated analysis code fails to execute, the repair loop retries, and if the retry also fails the error and the generated code are returned to the scientist rather than silently swallowed. Failures that cross into HPC are reported with a provenance link, so a scientist can see how far a plan progressed before it stopped. Failures of the provenance service itself are the least graceful. A provenance question simply cannot be answered while the service is unavailable, which is an argument for treating provenance as platform infrastructure with a corresponding availability target rather than as a per-project addition.

All model inference crosses MAG, so an outage of the gateway suspends the conversational layer. The offline pipeline involves no language model and continues unaffected, so the system degrades to a data-producing pipeline rather than failing outright. Similarly, the availability check depends on data published to the lakehouse: if publication lags, questions that would have been answered in seconds fall back to the cold path and take minutes.

\section{Generalizing to Other Use Cases}
\label{sec:general}

The point of describing OPAL in this detail is to make visible how little of it is about plants. Of the components in Fig.~\ref{fig:arch}, only the thick-bordered boxes are use case specific: a conversational agent, the compute agent logic, the APPL analysis server, and the domain model and trait algorithms. Everything else was inherited.

A use case adopting these capabilities inherits a place to run services, a versioned path for getting models onto remote compute, a governed inference endpoint, a substrate for publishing derived data, and provenance. It must supply three things. The first is a \emph{domain model and the code that derives quantities of interest from raw instrument data}; in OPAL this is the segmentation transformer and trait extraction, and it is irreducibly domain specific. The second is a \emph{schema and vocabulary} for its data, since the lakehouse stores what it is given and the agent's ability to generate correct analysis code rests on knowing what the columns mean; the trait synonyms in Section~\ref{sec:workflows} are an instance of this. The third is \emph{the decision logic of its agents}, namely what constitutes a complete analysis plan in the domain and when a computation is worth its cost.

Two structural patterns transfer with the capabilities and are, in our experience, where most of the engineering benefit lies. One is the \emph{availability check}: a use case that publishes derived products to the lakehouse can serve most requests from them and fall back to computation only when necessary, which is what produces the latency distribution in Table~\ref{tab:perf}. The other is \emph{cache-with-fallback model retrieval}: consulting the catalog only on a cold node keeps the steady-state path local while making the model version an explicit property of each run.

% Finally, the evaluation is a single use case at a single institution over one campaign. We report latency and overhead from that deployment rather than from a controlled comparison, and the manual baseline is the workflow the same group previously ran rather than an independently measured control. The generalization argument in Section~\ref{sec:general} is correspondingly structural rather than empirical.

\section{Related Work}

Several projects treat shared infrastructure as the unit of reuse for cross-facility science. The Superfacility model~\cite{superfacility} and work on cross-facility workflows~\cite{antypas,tyler} establish the operational patterns for coupling instruments with HPC, and INTERSECT~\cite{intersect} pursues an ecosystem of loosely coupled facility services. Work on secure programmatic facility access~\cite{s3m,etz} provides the substrate that makes cloud-to-HPC interaction tractable. Our contribution is complementary: rather than a facility integration framework, we report on a set of platform-operated capabilities and what a use case does and does not have to build on top of them.

Among systems that place language model assistants in front of facilities, VISION~\cite{vision} assembles models into task-specific cognitive blocks for operating synchrotron beamlines, and CALMS~\cite{calms} combines models with semantic search and tool execution to assist with experimental design and instrument operation, demonstrating that retrieval grounding materially reduces hallucination. Both target instrument operation; our use case begins after acquisition, where the bottleneck is interpretation, which shifts the emphasis toward HPC-scale analysis and its provenance. Our earlier work applied an agentic architecture to autonomous additive manufacturing at ORNL~\cite{rosendo}; the present work differs in framing the reusable substrate as platform capabilities rather than as a per-project architecture.

% On the individual capabilities, Academy~\cite{academy} provides abstractions for agents on federated infrastructure, Parsl~\cite{parsl} scales Python workloads on HPC schedulers, and Globus~\cite{globus} manages transfers between facility endpoints. For provenance, Flowcept~\cite{flowcept} captures multi-workflow provenance across the edge--cloud--HPC continuum and PROV-AGENT~\cite{provagent} extends the W3C PROV model~\cite{provdm} to agent decisions, with interactive agents evaluated as a means of making such records accessible in natural language~\cite{provllm}. Within plant phenotyping, toolchains such as PlantCV~\cite{plantcv} and deep learning approaches to trait prediction~\cite{deepplant} address extraction itself but assume a human drives the analysis.

\section{Conclusion and Future Work}

We reported on five Genesis platform capabilities and showed how they enable a cross-facility autonomous experiment. The demonstration is the OPAL plant phenotyping workflow, which couples ORNL's APPL facility with the Frontier supercomputer and lets biologists ask questions in natural language while a campaign is still running. We showed how the platform composes with external services it deliberately does not subsume and separated what the use case inherited from what it had to build: only the APPL analysis server, the conversational agent, the compute agent logic, and the domain model and trait algorithms were specific to plant science. In a 40-day nickel-treatment campaign the resulting workflow replaced roughly twelve hours of manual analysis with interactive queries returning in seconds to minutes, at under 1\% provenance overhead.

Two main directions follow. The first is completing the platform integration of the two capabilities that are still maturing. Flowcept is to become the Genesis platform provenance service, so that provenance is inherited by every use case rather than instrumented per project, with the resulting corpus curated as audit-grade training data for multimodal biology foundation models. The data lakehouse is in progress, and will require integration with the Genesis platform before other use cases can adopt it on the same terms that OPAL does today. The second is validating the capability boundaries against use cases with different shapes, particularly ones whose instrument produces streaming rather than batch data, and against deployments spanning multiple DOE laboratories.

\section*{Acknowledgment}
% This work, part of OPAL, is based upon work supported by the U.S. Department of Energy, Office of Science, through the Office of Biological and Environmental Research Program, under contracts DE-AC02-06CH11357 (ANL); DE-AC02-05CH11231 (LBNL); DE-AC05-00OR22725 (ORNL); and DE-AC05-76RL01830 (PNNL). This research used OLCF and APPL resources at ORNL, supported by the Office of Science of the U.S. Department of Energy under Contract No. DE-AC05-00OR22725. This work was performed under the DOE Genesis Mission.

This material by the Orchestrated Platform for Autonomous Laboratories (OPAL) is based upon work supported by the U.S. Department of Energy, Office of Science, through the Office of Biological and Environmental Research Program, under contracts DE-AC02-06CH11357 (ANL); DE-AC02-05CH11231 (LBNL); DE-AC05-00OR22725 (ORNL); and DE-AC05-76RL01830 (PNNL).This work supported by the U.S. Department of Energy, Office of Science, Office of Biological and Environmental Research under Award Number DE-SC-ERKPA89. This research used resources of the OLCF and the Advanced Plant Phenotyping Laboratory at ORNL, supported by the Office of Science of the U.S. Department of Energy under Contract No. DE-AC05-00OR22725.

\bibliographystyle{IEEEtran}
% Grouped so the references keep the \footnotesize the manual list used.
{\footnotesize
\bibliography{references}}

\end{document}